\documentclass[twocolumn,showpacs,floatfix,preprintnumbers,prl]{revtex4}
\usepackage{amssymb}
\usepackage{graphicx}
\usepackage{dcolumn}
\usepackage{bm}
\usepackage[usenames]{color}

\begin{document}

\title{Optical conductivity of doped Mott insulator: the interplay between 
correlation and electron-phonon interaction}
\author{G. De Filippis$^1$, V. Cataudella$^1$, A. S. Mishchenko$^{2,3}$, C. A. Perroni$^1$ and 
N. Nagaosa$^{2,4}$}
\affiliation{$^1$ CNR-INFM Coherentia and Dip. di Scienze Fisiche - Universit\`{a} di Napoli
Federico II - I-80126 Napoli, Italy\\
$^2$ Cross-Correlated Materials Research Group (CMRG), ASI, RIKEN,
Wako 351-0198, Japan
\\
$^3$RRC ``Kurchatov Institute'' - 123182 - Moscow - Russia\\
$^4$Department of Applied Physics, The University of Tokyo, 
7-3-1 Hongo, Bunkyo-ku, Tokyo 113, Japan}
\date{\today}

\begin{abstract}
The optical conductivity (OC) of cuprates is studied theoretically 
in the low density limit of the t-t$'$-J-Holstein model.
By developing a limited phonon basis exact diagonalization 
(LPBED) method capable of treating the lattice of  
largest size $4 \times 4$ ever considered,
we are able to discern fine features of the mid-infrared (MIR) 
part of the OC revealing three-peak structure. 
The two lowest peaks are observed in experiments 
and the highest one is tacitly resolved in moderately doped cuprates. 
Comparison of OC with the results of semianalytic approaches and 
detailed analysis of the calculated isotope effect  
indicate that the middle-energy MIR peak is of mostly magnetic 
origin while the lowest MIR band originates from the scattering of 
holes by phonons. 
\end{abstract}

\maketitle

The way to disclose the nature of the high temperature
superconductors lies on the understanding of the 
dynamics of the holes doped into a Mott insulator \cite{Lee}. 
It is recognized that the dynamics of holes is governed 
by the interaction with magnetic subsystem 
as it was proved by the angle resolved 
photoemission spectroscopy of 
underdoped compounds \cite{Shen_03}.
There is also a growing number of evidences 
that a considerable coupling to lattice  
contributes to the properties of the holes too 
\cite{Gunnar}.
These major interactions are expected 
to leave fingerprints in the the OC of cuprates. 
However, the interpretation of even
the basic features of the OC is controversial.

Not to say about fine structure, there is no agreement on the 
issue of how many peaks are seen in the OC, both theoretically and 
experimentally. 
Initially, only the Drude term and MIR peak at around 0.5 eV have 
been considered as contributions coming from the 
dynamics of charged carriers \cite{Dago}. 
Later, improved quality of the samples and experimental 
techniques gave an indication 
\cite{Thom91,Uchida91,Lupi92,Quij99,Waku04},
and finally clearly showed \cite{andrei1},
that there is at least one more band (MIR$_{\mbox{\scriptsize LOW}}$ band)
induced by doping in the energy range, 
$\sim 0.1$eV, which is just above the phonons energy.
Moreover the analysis of the 
experimental data suggests that also another 
contribution, peaked at 1.5eV, should be also 
taken into account
\cite{Uchida91,Lupi92,Quij99,Waku04}. 
Although there is a temptation to explain the later 
contribution as reminiscent of the charge-transfer 
peak in doped system, this third high energy peak
(MIR$_{\mbox{\scriptsize HIGH}}$ band) is observed
in La$_{2-x}$Sr$_{x}$CuO$_4$ at 1.5 eV which is 
considerably smaller than the peak energy at 2 eV, 
observed in undoped  compound \cite{Uchida91}.     
 
Even if the existence of a peak structure is recognized, its
nature has been debated.
Inability of the prototypical t-J model, where hole
moves in an antiferromagnetic background, to 
explain the experimental structure of OC attracted 
a significant interest on this problem. One possible direction consists in
considering the Hubbard model with moderate $U$.
The interpretation of the mid-IR peak,
based on a purely electronic effect (associated with the upper/lower-Hubbad
bands and some in-gap states induced by 
doping), has been proposed \cite{Basov}.
Recently it was concluded \cite{Millis}
that the moderate $U$ can reproduce the 
OC spectra in  La$_{2-x}$Sr$_{x}$CuO$_4$  
by $U \cong  4 eV$, which is in sharp contrast to 
the t-J picture in the strong correlation limit.
The other possible direction is to consider 
the additional electron-phonon interaction (EPI), 
i.e., t-J-Holstein (t-J-H) model, where hole interacts also
with dispersionless phonons. 
The OC of the latter model was calculated by several methods: 
exact diagonalization (ED) on the small 
$\sqrt{10}\times\sqrt{10}$ system \cite{Fehske}, 
Self-Consistent Born Approximation (SCBA) with respect 
to both phonons and magnons \cite{Mukhin}, 
Dynamical Mean-Field Theory (DMFT) for 
infinite dimensions \cite{Cappell}, Diagrammatic 
Monte Carlo (DMC) with SCBA for magnons \cite{andrei1}, 
and ED within the Limited 
Functional Space (EDLFS) \cite{Bonca}.
The fine structure of the OC in realistic 2D systems
can be studied only by DMC \cite{andrei1} and EDLFS 
\cite{Bonca} whereas the rest
of approaches encounter severe problems \cite{problem}. 
In the light of this statement it is extremely alarming 
that interpretation of the low energy peak of OC 
(POC$_{\mbox{\scriptsize LOW}}$) in \cite{andrei1} and 
\cite{Bonca} is different.
Magnetic origin of POC$_{\mbox{\scriptsize LOW}}$ 
is concluded in \cite{Bonca} while the statement about
phononic origin in \cite{andrei1} might be an error
originated from the SCBA in magnetic 
channel or spin wave approximation used in \cite{andrei1}.  
Therefore, the convincing evidence for the 
origin of the POC is an urgent and important
issue towards the understanding of the basic interactions
governing high temperature superconductors.

In the present Letter we study theoretically
the OC of t-t'-J-H model directly compared with
the experimental observations
supporting the vital role of EPI 
in cuprates. 
In addition to the inclusion of the realistic 
next nearest neighbor hopping
$t'$ to reproduce the oberved Fermi surfaces 
in cuprates \cite{Lee}, we avoid the spin wave and 
self-consistent Born approximations 
for the coupling to the spin system. 
By developing a LPBED method, we can calculate 
the OC of the largest ever considered 
$4 \times 4$ system.  
Due to the exponential growth of the basis
with size of the system, the $4 \times 4$ lattice
has considerably denser quantum states
than the $\sqrt{10}\times\sqrt{10}$ system
so that it is possible to resolve fine structure 
of the OC. 
For the first time we observe, in different ranges of EPI,
three peaks in OC, the highest one being seen because the spin-wave 
approximation is avoided.
Calculating the isotope effect, which induces
changes both in phonon frequency 
and exchange constant $J$, we show that, 
in the weak coupling regime, the low energy       
POC$_{\mbox{\scriptsize LOW}}$ and the middle energy 
POC are of phononic and magnetic origin, respectively. 
Furthermore, in the intermediate coupling regime the low energy 
POC$_{\mbox{\scriptsize LOW}}$ is still of purely 
phononic origin while the middle energy POC is a
mixture of the lattice and magnetic excitations. 
Finally, comparison with the results of different approximate 
schemes shows that the highest energy   
POC$_{\mbox{\scriptsize HIGH}}$ peak is due to incoherent 
transitions into the states unaffected by lattice 
deformation associated with the hole.  

The Hamiltonian for t-t$'$-J-Holstein model is a sum of 
t-t$'$-J Hamiltonian
\begin{eqnarray}
H_{tt^{'}J}=&& -t \sum_{i,\delta,\sigma} 
c_{i+\delta,\sigma}^{\dagger}c_{i,\sigma}
-t' \sum_{i,\delta^{'},\sigma} 
c_{i+\delta^{'},\sigma}^{\dagger}c_{i,\sigma} \nonumber \\
&& + \frac {J}{2} \sum_{i,\delta} S_{i+\delta} S_{i} 
- \frac {J}{8} \sum_{i,\delta} n_{i+\delta} n_{i} \; ,
\end{eqnarray}
EPI Hamiltonian
$
H_{h-ph} = \omega_0  g \sum_{i} (1-n_i) 
\left( a_{i}^{\dagger}+a_{i} \right) \;
$,
and dispersionless phonons Hamiltonian
$H_{ph}=\omega_0 \sum_i a_i^{\dagger }a_i$ 
($a_i^{\dagger }$ is the creation operator of a 
phonon at site i with frequency $\omega_0$).
Here $t$ represents the hopping amplitude of the site $i$ to nearest
neighbors $i+\delta$, $t'$ is the diagonal 
hopping amplitude to next nearest neighbors $i+\delta^{'}$,
$J$ is the exchange constant of 
the spin-spin interaction,
$c_{i, \sigma}$ is the fermionic operator with  
excluded double occupancy, $S_i$ is the 
$\frac {1}{2}$-spin operator at site $i$, and 
$n_i$ is the site $i$ number operator. 
We introduce the EPI dimensionless coupling constant, $\lambda=g^2 \omega_0/4 t$,
with the value $\lambda=1$ dividing the weak and strong coupling regimes
of the Holstein model in the adiabatic limit. 
Below we set $\hbar=1$, $t=1$, $J=0.3$, $t'=-0.25$, $\omega=0.15$
and OC is in units of $2 \pi e^2$. 
The one-hole ground state of the t-J model on $4 \times 4$ lattice
is sixfold degenerate.
This degeneracy between $(\pm \pi/2, \pm \pi/2)$, $(0,\pi)$ 
and $(\pi,0)$ is partially removed by $t'$ \cite{gagliano} providing a four-fold degenerate
ground state at momentum $(\pm \pi/2, \pm \pi/2)$.
Hence, one can naively expect that the OC should be sensitive 
to the value of t$'$.

The LPBED method is based on the modified Lanczos 
algorithm \cite{dagotto}, where 
magnetic degrees of freedom are treated exactly whereas the 
phonon variables are efficiently limited to a set which, as
it is shown below, gives better results than both the 
Momentum Average (MA) approximation \cite{berciu} and SCBA.   
We use the translational symmetry associated to 
periodic boundary conditions, 
requiring that the states have a definite momentum,  
and work in the one-hole subspace with $\sum_i S_i^z=\frac{1}{2}$. 
Each basis vector is a linear superposition with 
appropriate phases of the 16 translational copies of a state having 
a given hole location with assigned locations of the phonon quanta 
and spin flips (hole, spin and lattice configurations are 
together rigidly translated). 
All 6435 spin configurations of $4 \times 4$ lattice are included. 

The real bottleneck comes from Hilbert space required by the phonons
basis. 
For instance, if all the phonon configurations up to M=15 phonon states are included
the size of the system is strongly limited (    
$\sqrt{10}\times\sqrt{10}$ \cite{Fehske}). 
To circumvent this difficulty, LPBED keeps only two groups of
phonon states. 
The first group are the lattice configurations involving only
{\em single site} deformations, all the others being undeformed
$
\left | ph \right\rangle_{j}^{(n)} = 
(a_{j}^{\dagger})^n \left | 0 \right\rangle_j 
[\sqrt{n!}]^{-1} 
\prod_{i \ne j}\left| 0\right\rangle_{i}
$.
Here $j=1,.. N$ denotes the lattice sites, 
$\left | 0 \right\rangle_i$ is the i-site phonon vacuum state, 
and all possible $n=0,1,...M$ values are limited by $M=20$,
which is shown to be enough for convergence even in the
strong coupling regime \cite{Fehske}. 
Restriction to such basis makes the method equivalent 
\cite{barisic} to the MA approximation \cite{berciu}.
This is already a reasonable approximation, which satisfies exactly 
the first six spectral weight sum rules for the Lehmann spectral
function \cite{berciu}. 
To improve the MA scheme, LPBED method includes additional phonon states
with up to three ($n_1+n_2+n_3 \le 3$) phonons on {\em different lattice} sites
$
\left | ph \right\rangle_{j_1,j_2,j_3}^{(n_1,n_2,n_3)} = 
\prod_{h=1}^{3} 
(a_{j_h}^{\dagger})^{n_h} [\sqrt{n_h!}]^{-1}  
\left | 0 \right\rangle_{j_h}
\prod_{i \ne j_1, j_2, j_3}\left| 0\right\rangle_{i} \; .
$
Such limit is chosen since three-phonon basis is able to recover the SCBA 
contribution (up to three phonons) and it goes beyond including
all other processes of the same order which are not present in the 
SCBA approach \cite{Ramsak}. 
Hence, LPBED approach is better then MA and SCBA methods 
and the only domain where it can fail is the case when
there is a strong non-local deformation.
However, strong deformations are realised only in the 
strong coupling regime where deformation in the case 
of local Holstein interaction is just restricted to the 
hole position. 
Our approach, able to improve MA, is related to that introduced 
in Ref.~\cite{MonaGle07}, with the advantage that in 
LPBED method it is possible to calculate not only self energy
of the quasiparticle but any correlation function.  

At $T=0$ the OC at nonzero frequency is calculated using the 
Kubo expression of the OC in terms of the current-current 
correlation function  
$
\sigma_{xx}(\omega) = - (N \omega)^{-1} \Im 
\left( \Pi(\omega+i\eta) + \Pi(\omega-i\eta) \right)
$
with 
$
\Pi(\omega+i\eta) = \left\langle \psi_0 \right | j_x 
[\omega+i\eta-H+E_0]^{-1} j_x
\left | \psi_0 \right\rangle
$.
Here $\left | \psi_0 \right\rangle$ is the ground state 
($\vec{k}=(\frac {\pi} {2}, \frac {\pi} {2})$) with  
energy $E_0$, $N$ indicates the number of lattice sites, $\eta$ 
is a broadening factor that shifts the poles of $\sigma_{xx}(\omega)$ in 
the complex plane by replacing the
$\delta$ functions by Lorentians, and 
$
j_x = i e t \sum_{i,\delta,\sigma} 
c_{i+\delta,\sigma}^ {\dagger}c_{i,\sigma} (\vec{\delta})_x + 
i e t' 
\sum_{i,\delta^{'},\sigma} c_{i+\delta^{'},\sigma}^ {\dagger}c_{i,\sigma} (\vec{\delta^{'}})_x
$.

\begin{figure}
\flushleft
\includegraphics[scale=0.85]{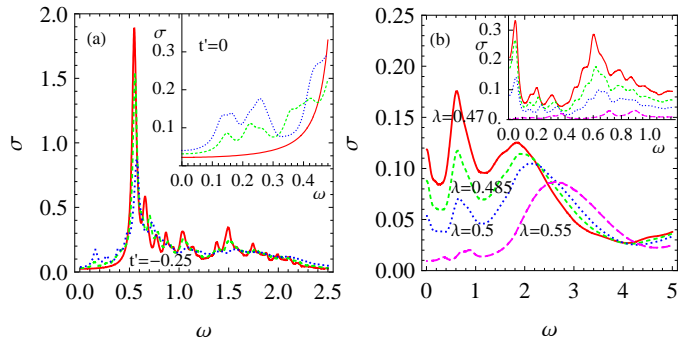}
\caption{(Color online)
OC in (a) weak [$\eta=0.025$] coupling regime
for $t'=-0.25$ and (inset) for $t'=0$.
OC at $t'=-0.25$ (b) from intermediate to strong 
coupling regime for $\eta=0.1$ and 
(inset) $\eta=0.025$.}
\label{fig:fig1}
\end{figure}

\begin{figure}
\flushleft
        \includegraphics[scale=0.85]{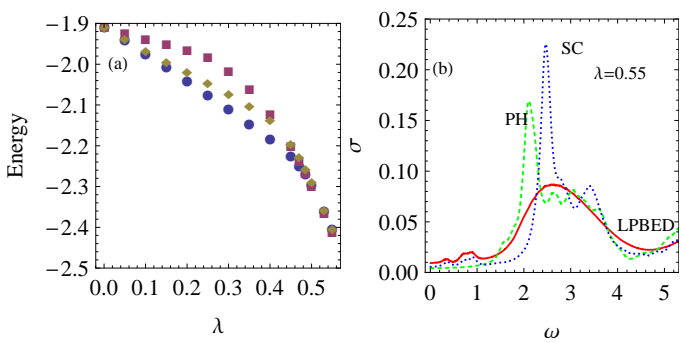}
        \caption{(Color online)
(a) Binding energy obtained within
LPBED method (circles), MA method (diamonds) and 
SC scheme (squares). (b) OC  
in the strong coupling regime ($\eta=0.1$).}
\label{fig:fig2} 
\end{figure}

For $\lambda=0$ we duly observe the well known  
POC in the t-J model 
at the energy 2J: the presence of  
$t'$ has little influence on the main feature 
of OC of t-t$'$-J model (Fig.~\ref{fig:fig1}a).
At weak EPI, in agreement with \cite{andrei1}, we detect 
POC$_{\mbox{\scriptsize LOW}}$ just above the phonon frequency. 
We stress that the existence of POC$_{\mbox{\scriptsize LOW}}$ is not related
to $t'$ hopping term, since the POC$_{\mbox{\scriptsize LOW}}$ is
also observed in the t-J-H model (inset in Fig.~\ref{fig:fig1}a).
This low energy peak appears only at nonzero EPI and persists up 
to the strong coupling regime (inset in Fig.~\ref{fig:fig1}b).
By increasing $\lambda$, the POC peak, which is around $2J$ at
$\lambda=0$, shifts to higher 
energies and its weight is gradually transferred to the high energy 
POC$_{\mbox{\scriptsize HIGH}}$ above $\omega\ge 2t$. 
Hence, the EPI changes the spectrum of the t-J model: OC exhibits 
three peaks. 
The nature of these peaks is either unknown or under dispute 
(cf.~\cite{andrei1} and \cite{Bonca}).
In the following we present several results unambiguously
revealing genesis of these peaks in the weak and strong 
coupling regimes. 

To study the origin of the three peaks for large EPI, 
we introduce strong coupling (SC) adiabatic approach in which
wave function is factorized into a product of 
normalized variational functions
$\left | \psi ({\bf r})\right\rangle$ and 
$\left | \phi ({\bf R})\right\rangle$ depending on 
electron ${\bf r}$ and phonon ${\bf R}$ 
coordinates, respectively.  
The expectation value of the Hamiltonian on the state 
$\left | \phi \right\rangle$ provides a Hamiltonian, $H_{el}$, depending 
only on the electronic degrees of freedom. It describes the t-t$'$-J model in 
a potential well. 
In Fig.~\ref{fig:fig2}a we compared results of SC approach 
for ground state energies of t-t$'$-J model with data obtained by 
LPBED method. 
For $\lambda$ above the critical $\lambda_c \approx 0.5$ the 
results of SC, MA, and LPBED approaches 
are in good agreement and, thus, the system is in the
strong coupling regime.  
In the SC limit, according to Franck-Condon principle, 
the lattice is frozen in the ground state during the hole optical excitations and the OC can be calculated considering 
excitations of the hole in the static potential well 
formed by the lattice deformation.
Comparison of SC result for OC with that obtained by LPBED for 
$\lambda=0.55$ (Fig.~\ref{fig:fig2}b) shows that SC approach 
reproduces all three peaks.   
We note that in the SC approach both initial and final electronic 
states are calculated in the lattice potential 
associated with the ground state wave function of the hole. 
On the other hand, if one assumes that the hole in the final state 
releases the deformation, we obtain another curve for OC with 
only the high energy POC$_{\mbox{\scriptsize HIGH}}$.
Such approach is often called "photoemission" (PH) process because 
the lattice deformation is only present in one of the counterparts 
of the states linked by current operator.  
Comparing results from LPBED, SC, and PH
approaches we conclude that POC$_{\mbox{\scriptsize LOW}}$
and POC in the strong coupling regime
represents hole transitions between states within the self-consistent 
potential well generated by the phonons.     
To the contrary, POC$_{\mbox{\scriptsize HIGH}}$ is associated 
with transitions into states which are intact by the EPI driven
lattice deformation. 
Here we note that the association of the theoretical 
POC$_{\mbox{\scriptsize HIGH}}$ with experimental
MIR$_{\mbox{\scriptsize HIGH}}$ band should be done with care,
since both structures  are located at energies which are near the 
limit of applicability of the t-J model, where one is 
required to use methods which can handle 
the initial unreduced three-band Hubbard model
\cite{Koikegami}. We stress that the above discussed scenario  
is valid also in the more simple t-J-Holstein model. 

Another way to establish the nature of the peaks of OC is to 
study changes of OC with small variations of the phonon frequency
$\omega_0$ and/or exchange constant $J$. 
Such study does not serve solely as Gedanken experiment but 
establishes how the spectra will be altered by the isotope 
substitution (IS).   
First of all, changes of the oxygen from $^{16}$O to $^{18}$O induces  
modifications in the values of coupling constant $g$ and phonon frequency 
$\omega_0$. 
Phonon frequency $\omega_0=\sqrt{k/M}$ is expressed in terms of
$k$, which is the restoring force for length unit of the 
local oscillators, and $M$, which is the mass of oxygen atoms
surrounding the Cu ion in the CuO$_2$ plain. 
IS changes the values of $\omega_0$ and $g$ 
to $\omega_0^ {*}=\omega_0\sqrt{M/M^{*}}$ and  
$g^ {*}=g (M^{*}/M)^{1/4}$ with 
value of $\lambda$ independent on isotope.  
In particular, the relative shift of $\omega_0$ is about $6\%$. 
The second possible effect of IS  
is the decrease of the antiferromagnetic exchange constant 
$J$ in compounds with the apical oxygen \cite{zhao,zhao2},
driven by its vibrations out of plane.
With the IS the value of $J$ is reduced by about 
$1\%$ and in the following we assume $\Delta J/J=-0.01$.  

\begin{figure}
\flushleft
        \includegraphics[scale=0.85]{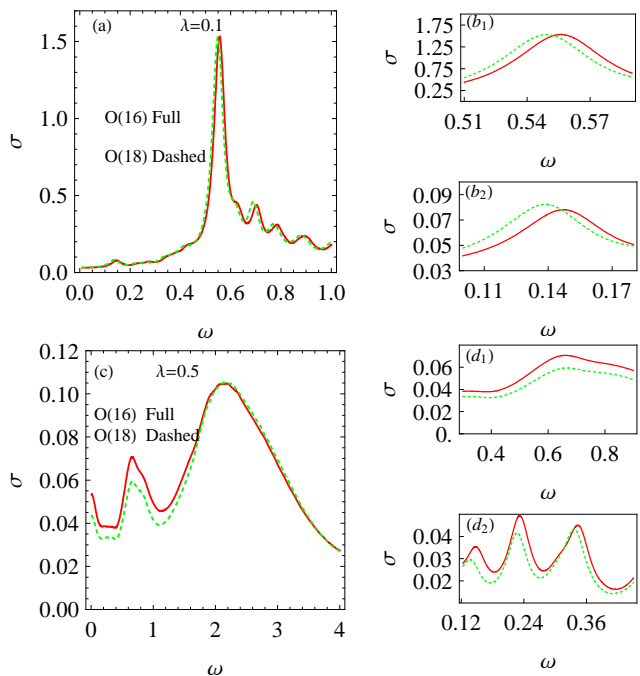}
        \caption{(Color online)
Effect of the O IS on OC in the weak 
[(a), (b$_1$), and (b$_2$)], and the intermediate coupling regime 
[(c), (d$_1$), and (d$_2$)].}
\label{fig:fig3}
\end{figure}

In Fig.3 we present the changes of OC induced both by change 
of the in-plane oxygen mass and exchange constant $J$. 
The contribution of these changes are well distinct with only
one exception.  
In the weak coupling regime (Fig.~\ref{fig:fig3}a and
Fig.~\ref{fig:fig3}b$_{1}$)  POC at frequencies around $2J$ is 
shifted down by about $1\%$ indicating the clear magnetic 
origin of this peak in the weak coupling regime. 
To the contrary, POC$_{\mbox{\scriptsize LOW}}$, 
with frequency around $\omega_0$, is shifted down about 6\% (Fig.~\ref{fig:fig3}b$_{2}$),
that is just softening of $\omega_0$ induced by IS.
Thus, we get one more confirmation of the 
phononic origin of theoretical 
POC$_{\mbox{\scriptsize LOW}}$ and experimental
MIR$_{\mbox{\scriptsize LOW}}$ band, in agreement with
Ref.~\cite{andrei1}.  

One gets the same conclusion about POC$_{\mbox{\scriptsize LOW}}$ 
from the IS in the intermediate 
coupling regime 
(Fig.~\ref{fig:fig3}c and Fig.~\ref{fig:fig3}d). Indeed 
the POC$_{\mbox{\scriptsize LOW}}$ is shifted down by 
6\% again. 
On the contrary, behavior of the middle energy POC
differs from that in the weak coupling regime
since it reduces its intensity and almost 
does not move with the IS. 
It is known that the middle POC energy increases by increasing
$\lambda/\lambda_c$ \cite{andrei1}. 
Moreover, $\lambda_c$ decreases by decreasing $\omega_0$ 
\cite{impurity}. 
Thus, the decrease of $\omega_0$ increases the energy of POC. 
To the contrary, decrease of $J$ tends to soften the POC. 
Hence, joint influence of both effects leaves the peak
at the same position. 
To confirm this conclusion we repeated the calculations setting to zero 
the change of $J$ and found (not shown) the increase
of POC energy. 
Hence, we conclude that in the intermediate coupling regime 
the middle POC is of mixed origin. 

In conclusion, we developed a novel method capable of 
studying the fine structure of the OC of the 
t-t$'$-J-Holstein model and found that the influence 
of the diagonal hopping t$'$ on OC is surprisingly little. 
For the first time we revealed a 
3-peak structure and established the origin of these
peaks in the whole range of hole-phonon couplings. 
We also predicted the influence of the IS on the 
OC which can be compared with experiment after the 
problem of residual $^{16}$O in the matrix of $^{18}$O,
strongly influencing low energy part of 
OC \cite{Bernhard}, will be solved.

A.S.M.\ was supported by RFBR 07-02-00067a, 
N.N. by Grant-in-Aids No.\ 15104006, No.\ 16076205, 
No. 17105002, No.\ 19048015, and NAREGI Japan; G.D.F., V.C. and C.A.P.
received financial support from Research Program MIUR-PRIN 2007.


\begin{thebibliography}{99}
%
\bibitem{Lee} P. A. Lee, N. Nagaosa and X.-G. Wen,
             Rev. Mod. Phys. \textbf{77}, 721 (2005).
%
\bibitem{Shen_03} A. Damascelli, Z.-X. Shen, and Z. Hussain, 
              Rev. Mod. Phys. {\bf 75}, 473 (2003).
%
\bibitem{Gunnar} O. Gunnarsson and O. R\"{o}sch,
             J. Phys.: Condens. Matter \textbf{20}, 043201 (2008).
%
\bibitem{Dago} E. Dagotto,
               Rev. Mod. Phys. \textbf{66}, 763 (1994). 
%
\bibitem{Thom91} G. A. Thomas et al., Phys. Rev. B \textbf{45}, 2474 (1992).
%
\bibitem{Uchida91} S. Uchida et al, 
                  Phys. Rev. B \textbf{43}, 7942 (1991).
%
\bibitem{Lupi92} S. Lupi et al
                 Phys. Rev. B \textbf{45}, 12470 (1992).  
%
\bibitem{Quij99} M. A. Quijada et al,
                Phys. Rev. B \textbf{60}, 14917 (1999).
%
\bibitem{Waku04} K. Waku et al,
                Phys. Rev. B \textbf{70}, 134501 (2004).
%
\bibitem{andrei1} A. S. Mishchenko et al,
          Phys. Rev. Lett., \textbf{100}, 166401 (2008).
%
\bibitem{Basov}
For a review see D.N. Basov and T. Timusk,
Rev. Mod. Phys. \textbf{77},721 (2005). 
%
\bibitem{Millis} A. Comanac et al, 
        Nat.\ Phys.\ \textbf{4} 287 (2008)
%
\bibitem{Fehske} B. Ba\"{u}ml, G. Wellein and H. Fehske,
                Phys. Rev. B \textbf{58}, 3663 (1998).
%
\bibitem{Mukhin} B. Kyung, S.I. Mukhin, V.N. Kostur and R.A. Ferrell,
                Phys. Rev. B \textbf{54}, 13167 (1996).
%
\bibitem{Cappell} E. Cappelluti, S. Ciuchi and S. Fratini,
                Phys. Rev. B \textbf{76}, 125111 (2007).
%
\bibitem{Bonca} L. Vidmar, J. Bonca and S. Maekawa,
                Phys. Rev. B \textbf{79}, 125120 (2009).
%
\bibitem{problem} DMFT introduces unrealistic infinite dimension, 
SCBA for phonons is unreliable \cite{tJpho}, and 
the spectrum of $\sqrt{10}\times\sqrt{10}$ in ED
is too sparse to see a fine structure.   
%
\bibitem{tJpho} A. S. Mishchenko and N. Nagaosa, 
                Phys. Rev. Lett. \textbf{93}, 036402 (2004).
%
\bibitem{gagliano} E. Gagliano, S. Bacci and E. Dagotto, Phys. Rev. B {\bf 42}, 6222 (1990).
%
\bibitem{dagotto} E. Dagotto and A. Moreo, Phys. Rev. D {\bf 31}, 865 (1985).
%
\bibitem{berciu} M. Berciu, Phys. Rev. Lett. {\bf 97}, 036402 (2006).
%
\bibitem{barisic} O. S. Barisic, Phys. Rev. Lett. {\bf 98}, 209701 (2007).
%
\bibitem{Ramsak} A. Ramsak and P. Horsch,
                 Phys. Rev. B \textbf{48}, 10559 (1993).
%
\bibitem{MonaGle07} M. Berciu and G.L.  Goodvin,
                    Phys. Rev. B \textbf{76}, 165109 (2007).
%
\bibitem{Koikegami} S. Koikegami and Y. Aiura, 
                   Phys. Rev. B \textbf{77}, 184519 (2008).
%
\bibitem{zhao} G.-M. Zhao, K. K. Singh, and D. E. Morris, Phys. Rev. B {\bf 50}, 4112 (1994).
%
\bibitem{zhao2} G.-M. Zhao, Phys. Rev. B {\bf 75}, 104511 (2007).
%
\bibitem{impurity} A. S. Mishchenko et al, 
                 Phys. Rev. B {\bf 79}, 180301(R) (2009). 
%
\bibitem{Bernhard} C. Bernhard et al, 
                  Phys. Rev. B \textbf{69} 052502 (2004).
\end{thebibliography}
\end{document}